\documentclass[fleqn,twoside]{article}
\usepackage{espcrc2}

\usepackage{graphicx}
\usepackage[figuresright]{rotating}
\usepackage{amsmath}
\usepackage{amssymb}

\newcommand{\sign}[0]{\text{sign}}
\newcommand{\bra}[1]{\left\langle #1\right|}
\newcommand{\ket}[1]{\left|#1\right\rangle}

\title{Current status of Dynamical Overlap project}

\author{N. Cundy
\address[WUP]{Theoretische Physik, Universit\"at Wuppertal, Gaussstrasse 19,
  D42109 Wuppertal}}
\begin{document}
\begin{abstract}
We discuss the adaptation of the Hybrid Monte Carlo algorithm to overlap
fermions. We derive a method which can be used to account for the delta
function in the fermionic force caused by the differential of the sign
function. We discuss the algoritmic difficulties that have been overcome, and
mention those that still need to be solved. 

\vspace{1pc}
\end{abstract}

\maketitle

\section{INTRODUCTION}
The overlap operator~\cite{neuberger} is the closest known lattice Dirac
operator to the continuum operator. With the lattice community currently moving beyond
the quenched approximation, one obvious possibility is to use dynamical
overlap fermions. The advantages of the overlap operator are well known: it
satisfies the Ginspag-Wilson lattice chiral symmetry exactly; there is an easy
non-perturbative renormalisation; there is no operator mixing
involving different chiral sectors; there is a well defined index ($Q_f=\frac{1}{2}\text{Tr}\epsilon(Q)$), equal
to the topological charge in the continuum limit; the anomaly is correctly
accounted for; and it is essential for stuties of topics such as topology, spontaneous
chiral symmetry breaking, and the eigenvalue spectrum of the Dirac operator. Since
chiral symmetry is so important to low energy QCD, it seems a waste not to use
the only Dirac operator which we know of which fully respects this symmetry.

Of course, there are reasons not to use the overlap operator. Firstly, it is
considerably slower than (for example) staggered or clover
fermions. Secondly, the discontinuity in the overlap operator creates a number
of unique problems when trying to impliment a Hybrid Monte Carlo
algorithm. The first of these issues will not be a problem once we have
sufficiently fast computers, which will be in the very near future (we have
already run some trajectories on a $16^3 32$ lattice). Now is the perfect time to tackle the
second problem, and to create a Hybrid Monte Carlo algorithm for overlap
fermions.

In this talk, I shall summerize the work done so far on this issue. Work has
been published in this area by Z. Fodor \textit{et al}~\cite{fodor1,fodor2,fodor3}, by myself in collaboration
with Thomas Lippert and Stefan Krieg~\cite{cundy1,cundy2,cundy3}, and by
T. DeGrand and S. Schaefer~\cite{degrand1,degrand2,degrand3}.

Section \ref{sec:two} provides a brief introduction to Hybrid Monte Carlo
and the overlap operator. Section \ref{sec:three} discusses the problem of
topological charge changes. Section \ref{sec:four} mentions some additional
problems and advantages concerned with dynamical overlap fermions. Section
\ref{sec:five} gives a few numerical results, and our conclusions are
presented in section \ref{sec:six}.
\section{HYBRID MONTE CARLO WITH THE OVERLAP OPERATOR}\label{sec:two}
The overlap Dirac operator is
  \begin{gather}
    D = (1+\mu) + \gamma_5(1-\mu) \epsilon(Q),
  \end{gather}
where $\mu$ is a mass parameter, $Q$ is the hermitian Wilson Dirac operator,
and $\epsilon$ is the matrix sign function. In our numerical simulations~\cite{cundy1}, we use a Zolotarev rational
approximation to the sign function~\cite{Zol} with the small eigenvalues treated exactly using
eigenvalue projection, but for the purposes
of this talk, I shall assume that we can calculate all the eigenvalues
$\lambda_i$ and eigenvectors $\bra{\psi_i}$ of $Q$
and thus treat the matrix sign function exactly using $\epsilon(Q)=\text{sign}(\lambda_i)\ket{\psi_i}\bra{\psi_i}$. This will simplify the
algebra while retaining the important features of the algorithm. We will define $H$ as the Hermitian
overlap operator $\gamma_5 D$.

The Hybrid Monte-Carlo (HMC) algorithm~\cite{HMC} updates the gauge field in two steps:
(1) a molecular dynamics (MD) evolution of the gauge field;
(2) a Metropolis step which renders the algorithm exact. In the MD step, we introduce a momentum, $\Pi$, which is conjugate to
the gauge fields $U$, and a spinor field $\phi$ which is used
to estimate the fermion determinant via a heat-bath. We define an Hybrid Monte-Carlo energy
\begin{gather}
E= \frac{1}{2}\Pi^2 + S_g[U] +\phi^{\dagger}(H)^{-2}\phi.
\end{gather}
$S_g$ is the gauge action, and we will use either the Luscher-Weisz or Wilson
plaquette action. We introduce a computer time $\tau$ and integrate over the classical equations
of motion to generate the correct ensemble. We cannot perform
an exact integration, so we need to use a numerical method such as the
Omelyan integration step~\cite{hep-lat/0505020}. This
will create a small error in the energy conservation, which we can correct for by
including an additional metropolis step, accepting or rejecting the
new configuration according to a probability $P_{\text{acc}}=\min(1,\exp(\Delta)$, where
$\Delta = E_i-E_f$, $E_i$
is the initial energy and $E_f$ the energy at the end of the MD. It is
therefore important that we conserve energy as well as possible during the
MD to ensure a high acceptance rate. Note that we do not have to use the
classical trajectory: any reversible update which leads to a small $\Delta$ will
suffice. Our MD procedure does not even have to conserve area, as long as we
can easily calculate the Jacobian $J$. If we have a non-area conserving
update, we just have to include the Jacobian in the Metropolis step, using a new 
$\Delta= E_i-E_f+\log J$~\cite{Borici}. To have a high acceptance rate, we need
$e^{\Delta}\sim 1$.

The crucial part of the MD procedure is the force used to update the momentum, defined as
\begin{gather}
F_T = -U \frac{\partial(S_g[U] +\phi^{\dagger}(D^{\dagger}
D)^{-1}\phi)}{\partial U}.
\end{gather} 
The fermionic part of this force for the overlap operator is
\begin{align}
&F_F\Pi+\Pi F_F^{\dagger}=-(1-\mu^2)\nonumber\\
&\phi \frac{1}{H^2} \left(\gamma_5\frac{d}{d\tau} \epsilon (Q) +
  \frac{d}{d\tau} \epsilon (Q)\gamma_5\right)\frac{1}{H^2}\phi
\end{align}
We can differentiate the eigenvectors and eigenvalues of $Q$ using a procedure
analogous to first order time independent perturbation theory in quantum
mechanics~\cite{cundy1}. This gives
\begin{align}
&\frac{d}{d\tau} \epsilon (Q) = \nonumber\\
& \sum_{i,j\neq i}
\ket{\psi_i}\bra{\psi_i}\frac{dQ}{d\tau}\ket{\psi_j}\bra{\psi_j}\frac{\sign(\lambda_i)-\sign(\lambda_j)}{\lambda_j
  - \lambda_i}\nonumber\\
& +\sum_i \ket{\psi_i}\bra{\psi_i} {\frac{d}{d\tau} \sign(\lambda_i)}.\label{eq:force}
\end{align}
Note that only mixings between eigenvalues of different signs contribute to
the fermionic force, and only mixings between the small eigenvalues are
important.
 The main feature of the fermionic force is the
Dirac $\delta$-function coming from the differential of the sign function. We
shall discuss how to deal with this in the next section.
\section{EIGENVALUE CROSSINGS}\label{sec:three}
The $\delta$-function in the fermionic force, which occurs whenever one of
the Wilson eigenvalues crosses zero (i.e. whenever there is a change in the
index of the overlap operator), should, in an exact integration, introduce a
discontinuity in the momentum, which will exactly cancel the discontinuity in
the pseudo-fermion energy caused by the abrupt change in the matrix sign
function. We can visualise this in a classical mechanics picture by picturing a
potential wall of height $-2d$ surrounding each topological sector. We can
easily calculate the height of the wall (which can be either positive or
negative), either by integrating the fermionic force across the $\delta$-function, or by calculating the difference in the pseudo-fermion energy. Both
these procedures give
\begin{align}
&d =
-(1-\mu^2)\nonumber\\&\bra{\phi}\frac{1}{(H^+)^2}\left\{\gamma_5,\epsilon(\lambda^-)\ket{\psi}\bra{\psi}\right\}
&\frac{1}{(H^-)^2}\ket{\phi}.
\end{align} 
$H^+$ is the Hermitian overlap operator just after the crossing, $H^-$ the operator just
before the crossing.
In a classical mechanics picture, the momentum would be updated in a direction
parallel to $\eta$, the normal to the topological sector wall, thus:
$(\Pi^+,\eta)^2 =(\Pi^-,\eta)^2+4d$. If the momentum is too small (i.e. this
procedure can lead to an imaginary momentum) we reflect of the potential
wall. However, as remarked earlier, there is no
reason why we have to stick to the classical mechanics picture. Below, we
shall describe how a general updating procedure can be derived to account for
the potential wall. 

To calculate the Jacobian, we need to work with the coordinate and momentum vectors $u$ and
$\pi$. $\pi$ can be calculated easily from the momentum field
$\Pi_{\mu}(x)=T_i\pi^i_{\mu}(x)$, where $\mu$ and $x$ refer to the direction
and lattice site, and
$T_i$ are the generators of the gauge group. The gauge coordinate $u$
is defined so that an update of the gauge field $U\rightarrow e^{i\Pi}U$
corresponds to $u\rightarrow u+\pi$. We correct for the $\delta$-function in three steps. (1) We update the gauge
field to the potential wall $u_c=u^- +\tau_c \pi^-$; (2) We update the
momentum, using $(\pi^+_i)^2 = (\pi^-_i)^2 + G_i(\pi^-,u_c)$, where we shall
determine the functional form of $G$ later; (3) We return the
gauge field to the original point  $u_c=u^+ -\tau_c \pi^+$. Here $\tau_c$ is
the computer time at which the eigenvalue is zero, i.e. 
\begin{gather}
\tau_c=\frac{(u_c-u,\eta)}{(\pi^-,\eta)}.
\end{gather}
$(\pi^-,\eta)$ refers to the scalar product of the two vectors. Differentiating $\tau_c$ with respect to $u$ and $\pi$ gives
\begin{gather}
\frac{\partial \tau_c}{\partial \pi_k} = \tau_c\frac{\partial \tau_c}{\partial
  u_k} = -\tau_c\frac{\eta_k}{(\pi,\eta)}.
\end{gather}
Any function, $g$, of $u_c$ (such as $d$ or $\eta$) will obey the relation
\begin{gather}
\frac{\partial g}{\partial \pi_k} = \tau_c\frac{\partial g}{\partial
  u_k}
\end{gather}
We are now in a position to write down the Jacobian. Using the gauge update
above gives
\begin{align}
J =&\left|\begin{array}{c c}
\frac{\partial \pi^+_i}{\partial \pi^-_k}&\frac{\partial \pi^+_i}{\partial u^-_k}\\
\frac{\partial u^+_i}{\partial \pi^-_k}&\frac{\partial u^+_i}{\partial u^-_k}
\end{array}\right|\nonumber\\
\frac{\partial u^+_i}{\partial
  \pi^-_k}=&\tau_c\delta_{ik}+\frac{\partial{\tau_c}}{\partial
  \pi^-_k}(\pi^-_i-\pi^+_i) - \tau_c\frac{\partial \pi^+_i}{\partial
  \pi^-_k}\nonumber\\
\frac{\partial u^+_i}{\partial u^-_k}=&\delta_{ik} +
\frac{\partial{\tau_c}}{\partial u^-_k}(\pi^-_i-\pi^+_i) - \tau_c\frac{\partial \pi^+_i}{\partial u^-_k}
\end{align}
We can calculate $J$ using two quick determinant manipulations. We subtract
$\tau_c$ times the top row from the bottom row. We then subtract $\tau_c$
times the right column from the left column. These manipulations kill the
bottom left hand element. Two lines of algebra later, and we obtain
\begin{gather}
J=\left|
\left(\frac{\partial \pi^+_i}{\partial
  \pi^-_k}\right)_{u_c}\right|\frac{(\eta,\pi^+)}{(\eta,\pi^-)}.
\end{gather}
We consider the components of the momentum normal to $\eta$ and perpendicular to $\eta$
separately. For example, if we update the momentum normal to $\eta$, then we
use the momentum update above and insert this Jacobian into the condition
$e^{\Delta}=1$ needed for a high acceptance rate. This
immediately gives us a differential equation for $G_{\eta}$:
\begin{gather}
e^{-G_{\eta}/2+2d}\frac{1}{\pi^+}\left(\pi^-_{\eta}+\frac{1}{2}\frac{\partial G_{\eta}}{\partial
  (\pi^-_{\eta})}\right)\frac{\pi^+_{\eta}}{\pi^-_{\eta}}=1.
\end{gather}
It is trivial to solve this equation to obtain the momentum update
\begin{align}
&e^{-(\pi^+_{\eta})^2/2}-e^{-(\pi^-_{\eta})^2/2 -2d}\nonumber\\
&\phantom{spacespace}-A(|d|)(1-e^{-2d})=0.\label{eq:transmission}
\end{align}
We have written the constant of integration as $A(1-e^{-2d})$ to
ensure reversibility. $A$ should lie in the range $0\le A\le1$. $A=0$ gives us
the classical mechanics solution. 

We cannot allow the final momentum to be
complex. Therefore, if the initial momentum is in the range $a<\exp(-(\pi^-_{\eta})^2/2)<b$,
where
\begin{align}
a=&A\left(1-e^{2d}\right)\nonumber\\
b=&e^{2d}(1-A)+A,
\end{align}
we can use equation (\ref{eq:transmission}) to update the momentum (we call this case \textit{transmission}).  Following
~\cite{fodor1}, we reflect the momentum of the potential wall if the momentum
lies outside this range (i.e. we use $\pi^+_{\eta}=-\pi^-_{\eta}$, with some
additional terms -- outlined in ~\cite{cundy1} -- to ensure O($\tau^2$)
energy conservation). We need to keep the transmission rate as high as
possible in order to reduce the topological autocorrelation.
We can calculate the transmission rate by assuming that the momentum is initially
distributed according to $\exp(-(\pi^-_{\eta})^2/2)$. The probability of transmission is
\begin{gather}
P_t=\int_{\sqrt{-2\log(a)}}^{\sqrt{-2\log(b)}}e^{-(\pi^-_{\eta})^2/2}d\pi^-_{\eta}
\end{gather}
and we can show that $\partial P_t/\partial A>0$ for $0>A>1$. Therefore $A=1$
maximises $P_t$.

We can also update the momentum in directions perpendicular to $\eta$. The
procedure is exactly the same as above, and there are a large number of possible
solutions, depending on the number of dimensions and how we combine the gauge
fields. The simplest is, perhaps, the two dimensional case. We write
$\pi_1=r\cos\theta$ and $\pi_2=r\sin\theta$. We can now proceed to change
$r$\footnote{If we prefer, we can also change $\theta$, using $\theta\rightarrow
  e^{2d}\theta$.},
using
\begin{align}
&e^{-(r^+)^2/2}-e^{-(r^-)^2/2 -2d}\nonumber\\
&\phantom{spacespace}-A(|d|)(1-e^{-2d})=0.\label{eq:perpupdate}
\end{align}
We can also
update as many of these pairs of momentum fields as we like, or work in more
than two dimensions. However, no matter what update we use perpendicular to
$\eta$, the probability of transmission remains ($\min(1,e^{2d}$). However, we
can use the updates perpendicular to $\eta$ to remove one rather large
annoyance: an O($\tau_c$)
energy violating term. Without some means of removing this, we would have to
reduce the time step to unfeasibly small values: it would cripple the algorithm.

It is easy to show that the procedure outlined above only conserves energy up
to order $\tau_c$, because we update the gauge field, but not the momentum
field, to the topological sector wall. We can update the momentum field using the force
perpendicular to $\eta$ without affecting the Jacobian (as long as we only change
components of the momentum field orthogonal to this force), but we cannot so
easily use the component of the force normal to $\eta$. This leaves us with an
energy violation $\Delta E = \tau_c(F^+,\eta)(\eta,\Pi^+)-\tau_c(F^-,\eta)(\eta,\Pi^-)$. We can,
however, add
this term to the perpendicular update, for example by changing $d\rightarrow
d-\Delta E$ in equation (\ref{eq:perpupdate}). This allows us to remove the
O($\tau$) and many of the O($\tau^2$) energy violating terms,
leaving us with a correction step that is almost O($\tau^3$).
\section{OTHER ISSUES}\label{sec:four}
We had
\begin{align}
d =&
-(1-\mu^2)\bra{\phi}\frac{1}{(H^+)^2}\nonumber\\
&\left\{\gamma_5,\epsilon(\lambda^-)\ket{\psi}\bra{\psi}\right\}\frac{1}{(H^-)^2}\ket{\phi}.\nonumber
\end{align}
The good news is that this is (approximately) independent of the volume.
However, it is proportional to the inverse square of the mass $\mu$ (see table
\ref{tab:masses} for a rough numerical confirmation of this). The probability
of transmission has an exponential dependence on $d$. Therefore, at small
masses we are going to have a low acceptance rate. This can partially be
solved by introducing multiple pseudo-fermion fields~\cite{degrand3,Hasen},
but it still remains a serious issue that still needs to be resolved. We also
expect problems changing topological sector at small lattice spacing.

Zoltan Fodor and his collaborators have recently developed a novel algorithm to avoid the necessity
of changing topological sectors~\cite{fodor3}. It is easy, by
continually reflecting, to keep the simulation fixed within one topological
sector. By starting in several different topological sectors, it is possible
to calculate the expectation value of an observable with each topological
sector. To calculate the total expectation value, one has to find the
relative weighting of the various sectors. This can be done by measuring an
observable (which is defined only on the topological sector wall) on either side of the wall. Although there are still some doubts concerning the ergodicity of this
method\footnote{But see the discussion in section 3 of ~\cite{fodor3}.}, it does represent an interesting possibility to solve the problem of
the topological autocorrelation at small masses. 
  
\begin{table*}[htb]
\caption{The dependence of the potential wall on lattice size and mass
  $\mu$. The ensembles were generated using the Luscher-Weisz gauge action.}
\label{tab:masses}
\renewcommand{\tabcolsep}{2pc} 
\renewcommand{\arraystretch}{1.2} 
\begin{tabular}{l l l l l}
lattice size&$\beta$&$\mu$&$<d>$&$-<d>\mu^2$\\
\hline
$4^4$&7.5&$0.2$&$-2.21$&0.084\\
$12^4$&7.5&$0.1$&$-6.49$&0.065\\
$4^4$&7.5&$0.05$&$-35.92$&0.090
\end{tabular}
\end{table*}

Another problem is that the fermionic force can be unstable if we have two
small eigenvectors of opposite signs --- the force is proportional to
$1/(\lambda_1-\lambda_2)$ (see equation (\ref{eq:force})). This can be reduced by using stout
smearing~\cite{stout} or an improved kernel operator to reduce the number of
small eigenvalues~\cite{degrand2}, but these do not address the underlying problem. We shall discuss
this issue further in a future publication.

One significant advantage of overlap fermions is that the chiral symmetry allows us to
factorise the squared overlap operator into the two chiral sectors:
\begin{align}
&2+\gamma_5\epsilon(Q)+\epsilon(Q)\gamma_5 =\nonumber\\
&\left[\frac{1}{2}{(1+\gamma_5)} + \alpha\frac{1}{2}{(1-\gamma_5)}
    +\right.\nonumber\\
&\left.\phantom{spacespace}
  \frac{1}{4}{(1+\gamma_5)}\epsilon(Q){(1+\gamma_5)}\right]\times\nonumber\\
&\left[\frac{1}{2}{(1-\gamma_5)} + \alpha\frac{1}{2}{(1+\gamma_5)} -\right.\nonumber\\&\left.\phantom{spacespace}
  \frac{1}{4}{(1-\gamma_5)}\epsilon(Q){(1-\gamma_5)}\right]\frac{2}{\alpha}\nonumber
\end{align}
$\alpha\neq0$ is an arbitrary constant. Both of the factors are positive
definite, so we can use this decomposition to run single flavour
simulations. It is easy to show that these two operators have the same non-zero
eigenvalue spectrum as the overlap operator $H$ (up to an unimportant sign). Zero
modes can be included for by introducing additional pseudo-fermion fields to
generate the determinant of $1-\frac{1-\mu}{1+\mu}\epsilon(Q)$. This is only
useful at large masses, because we have to use a polynomial or rational
approximation to obtain the square root of the chiral projected overlap
operator, but we have tested it at approximately the strange quark mass on small lattices (using two single
flavour simulations) and the results for the plaquette and topological
susceptibility agree with the 2-flavour HMC. This method can
be used to simulate a 2+1 (or 2+1+1 etc.) flavour theory with little
additional effort.

An important question is how well this algorithm scales with the volume. It
has been suggested that the algorithm scales as the square of the volume $V$~\cite{degrand2}. The work needed to perform a correction step
is is proportional to the volume, and the density of small modes of the Wilson
operator is also proportional to the volume. If the number of crossings were
proportional to the density of small eigenvalues, then this would lead to an
O($V^2$) algorithm. However, it is by no means certain that this is the
case because small eigenvalues with opposite signs repel during the molecular
dynamics. Our numerical experience is that although the number of crossings
increases as we increase the volume, it scales considerably better than
O($V$). On small lattices we observed a $V^{1.5}$ scaling for the entire HMC algorithm, although this needs
to be checked on larger volumes. More work needs to be done on this area to
fully answer this important question.
\section{RESULTS}\label{sec:five}
\begin{figure*}[htb]
\vspace{9pt}
\begin{tabular}{c}
\includegraphics[width=15cm,height=6cm]{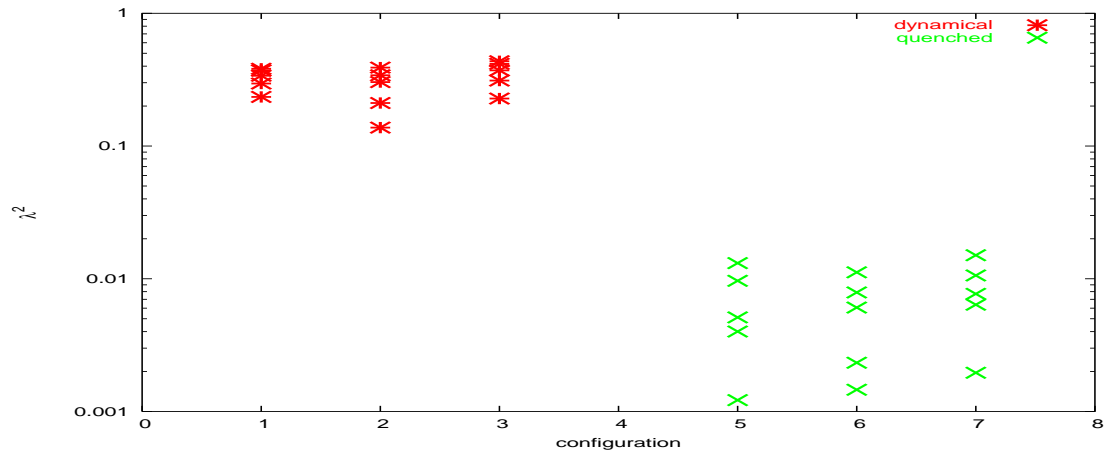}
\end{tabular}
\caption{The smallest non-zero eigenvalues for the overlap operator for a
  $\mu=0.1$ dynamical
  ensemble (left), and quenched ensemble (right) on $12^4$ lattices with
  lattice spacing $\sim 0.18 fm$.}\label{fig:eigenvalues}
\end{figure*}
\begin{figure*}[htb]
\vspace{9pt}
\begin{tabular}{c}
\includegraphics[width=15cm,height=6cm]{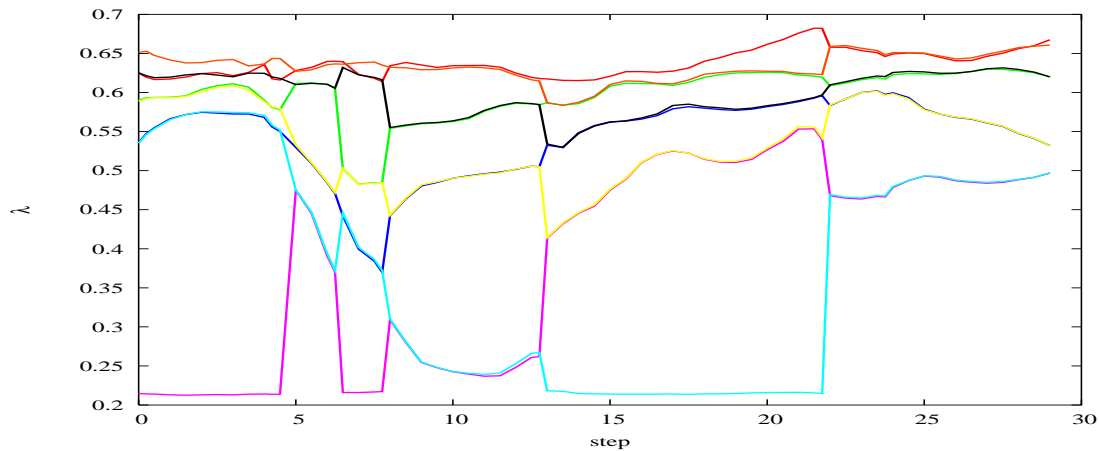}
\end{tabular}
\caption{The eight smallest eigenvalues of the overlap operator during a molecular dynamics trajectory.}\label{fig:eigenvalues2}
\end{figure*}

In figure \ref{fig:eigenvalues}, we compare the small eigenvalues of the
squared overlap operators for quenched and dynamical ensembles on $12^4$
lattices at (approximately) the same lattice spacing. The dynamical
configurations were generated at a mass $\mu=0.1$, although both sets of
eigenvalues were plotted with the massless overlap operator. It can be seen
that fermion determinant significantly suppresses the small eigenvalues of the
Dirac operator. This is, of course, to be expected, because small eigenvalues
would reduce the determinant. However, it is significant because we know from
the Banks-Casher relation that the small eigenvalues lead to the spontaneous
breaking of chiral symmetry. However, although the configurations shown in
figure \ref{fig:eigenvalues} do not posses small eigenvalues, we have seen
them during the MD.\footnote{We discontinued the run which we used to generate
figure \ref{fig:eigenvalues} shortly after the plot was made because the lattice
  spacing was too large. We do not yet have enough data on our current runs to
  generate an improved plot.} Figure \ref{fig:eigenvalues2} plots the
eigenvalues of the overlap operator during one molecular dynamics
trajectory. There were five topological charge changes during this
trajectory. The trajectory started with a topological charge -1, and the
eigenvalues of magnitude $\sim 0.21$ are the zero modes (calculated to about a
10\% accuracy with a mass $\mu=0.1$). The non-zero modes at either end of the
plot are similar in magnitude to those in figure \ref{fig:eigenvalues}, and
these will not be responsible for chiral symmetry breaking. There is a
considerably smaller
non-zero eigenvalue between the third and fourth topological charge
changes. A possible interpretation of this figure is that we create an
anti-instanton on the second topological charge change, and an instanton on
the third.\footnote{Although I am referring to instantons and anti-instantons,
  recent quenched calculations suggest that topological vacuum is not in fact
  dominated by instantons, but by long range topological fluctuations. But
  this argument holds however the zero modes are generated: I only use the
  terms instantons and anti-instantons for the sake of simplicity.} The small
eigenvalue is then be generated by the mixing between the two zero
modes. If this interpretation is correct, then this presents further evidence
for an underlying topological cause for chiral symmetry breaking.

\begin{figure*}[htb]
\vspace{9pt}
\begin{tabular}{c}
\includegraphics[width=15cm,height=6cm]{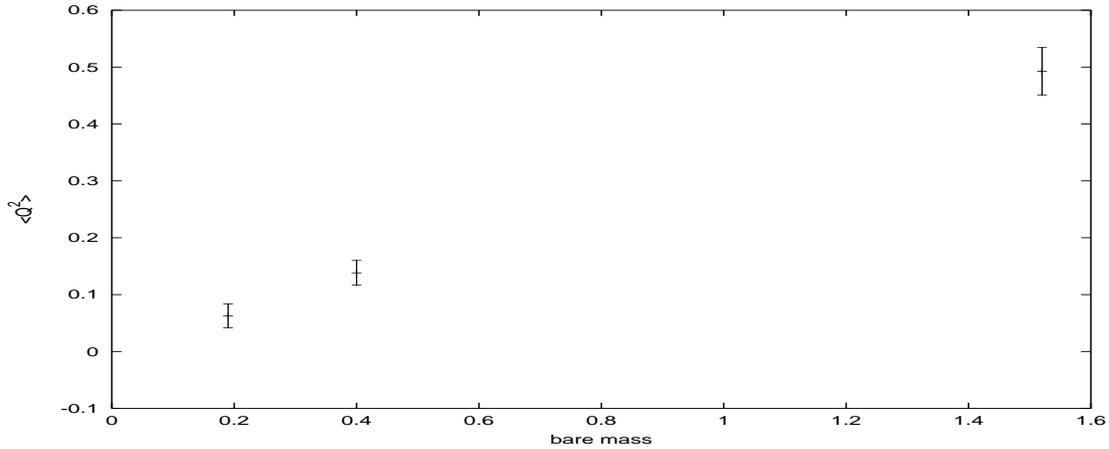}
\end{tabular}
\caption{The average value of the squared topological charge $Q_f^2$ on $6^4$ lattices plotted against quark
  mass with the Wilson
  plaquette action at $\beta=5.4$}\label{fig:topsup}
\end{figure*}

One area in which we expect dynamical overlap simulations to be particularly
important is the measurement of the topological susceptibility, defined as
$\chi_t=\langle Q_f^2\rangle/V$, where $Q_f$ is the index of the overlap
operator, and $V$ is the lattice volume, since overlap
fermions are the the only lattice fermions with a well defined index
theorem. In figure \ref{fig:topsup} we show plot the average value of the
squared topological charge $\langle Q_f^2\rangle$
against quark mass. It proved impossible to calculate the lattice spacing to
any accuracy on these small lattices, so we have no data for $\chi_t$
itself. The lattice spacing will change with the quark mass, but
not significantly. Although we are unable to calculate the volume, we
do see the expected linear dependence of the topological susceptibility with the quark mass.

\section{CONCLUSIONS}\label{sec:six}
Dynamical overlap fermions offer an exiting prospect for lattice QCD at small
masses. However, the HMC is considerably harder to implement
because of the discontinuity in the .Dirac operator The largest problem, how to deal
with the Dirac-delta function in the fermionic force, has been solved. We still have
a number of smaller issues to resolve, such as the problem of the topological
autocorrelation at small masses and due to mixings between small eigenvectors
of opposite signs. However, progress is being made on these issues. Although
overlap fermions are still rather slow, we hope to begin large-scale
simulations in the near future. We are currently working on lattices with
sizes up to $16^3 32$ and at masses of approximately one third of
the strange quark mass, and larger lattices and smaller quark masses should be
possible on the next generation of computers.

\section{Acknowledgements}
I would like to thank in particular my collaborators Th. Lippert and
 S. Krieg,
 and also A. Borici, T. DeGrand, G. Egri, Z. Fodor, T. Kennedy, S. Schaefer
 G. Schierholz, H. Steuben, T. Streuer and
 K. Szabo  for many useful
discussions. The computer simulations were carried out on the Wuppertal
AliceNext cluster, and the J\"ulich Blue-Gene, and I would like to thank
Norbert Eicker for his help administering these machines. NC was
supported by the EU grant MC-EIF-CT-2003-501467 and this work was part of the EU integrated infrastructure initiative HADRONPHYSICS
project under contract number RII3-CT-2004-506078 as well as the EU
integrated infrastructure project I3HP ``Computational Hadron
Physics'' contract No. RII3-CT-2004-506078.

\end{document}